\documentclass[prb,aps,floatfix,amsmath,amssymb,superscriptaddress,tightenlines,twocolumn]{revtex4}

\usepackage{epstopdf}
\newcommand{\be}{\begin{equation}}
\newcommand{\ee}{\end{equation}}
\usepackage{amsmath}
\usepackage{amsfonts}
\usepackage{bm}
\usepackage{color}
\usepackage{amsthm}
\usepackage[latin1]{inputenc}
\usepackage{graphicx}
\usepackage{subfigure}

\newcommand{\bra}[1]{\left\langle #1 \right|}
\newcommand{\ket}[1]{\left|#1\right\rangle}

\newcommand{\Tr}{\textrm{Tr}}

\begin{document}
\title{Entropic topological invariant for a gapped one-dimensional system}
\author{Isaac H. Kim}
\affiliation{Institute of Quantum Information and Matter, California Institute of Technology, Pasadena CA 91125, USA}

\date{\today}
\begin{abstract}
We propose an order parameter for a general one-dimensional gapped system with an open boundary condition. The order parameter can be computed from the ground state entanglement entropy of some regions near one of the boundaries. Hence, it is well-defined even in the presence of arbitrary interaction and disorder. We also show that it is invariant under a finite-depth local quantum circuit, suggesting its stability against an arbitrary local perturbation that does not close the energy gap. Further, it can unambiguously distinguish Majorana chain from a trivial chain under a global fermion parity conservation. We argue that the order parameter can be in principle measured in an optical lattice system.
\end{abstract}

\maketitle

\section{Introduction}
Topological order is a new kind of order that cannot be described by Landau's symmetry breaking paradigm. Systems with topological order are known to exhibit a number of intriguing properties which have no counterpart in classical physics. Examples include topological ground state degeneracy,\cite{Wen1992} emergent anyonic particle statistics,\cite{Kitaev2003,Levin2005} and topological entanglement entropy.\cite{Kitaev2006,Levin2006}

Kitaev's Majorana chain also has some of these properties.\cite{Kitaev2001} It can encode a qubit in its ground state that is robust against a weak perturbation that respects the fermion parity, and one can construct networks of such chains to braid Majorana fermions.\cite{Alicea2010} However, topological entanglement entropy - defined as a constant subcorrection term of the entanglement entropy - has not been discussed in this system to the best of author's knowledge.

There are several reasons to believe why such a quantity might be ill-defined, or even completely nonexistent. For a one-dimensional system, entanglement entropy satisfies an area law.\cite{Hastings2007a} Therefore, the subleading term must vanish asymptotically, which is unlikely to contain any useful information about the phase. 

However, on the other hand, the author has recently proved an inequality between topological entanglement entropy and topological ground state degeneracy.\cite{Kim2013} Since Majorana chain has two topologically protected ground states, one may expect some form of topological entanglement entropy to be present in the system. We explicitly show that such a speculation is indeed correct. Furthermore, we present a stability argument based on a well-known fact: that adiabatic evolution can be simulated by a finite-depth local quantum circuit.\cite{Hastings2004a,Hastings2005,Osborne2006}.

We note in passing that our framework is quite general, in that the only assumption we make about the Hamiltonian is its geometrically local structure and the existence of an energy gap that is independent of the system size. Nevertheless, we focus on the application to the Majorana chain in order to convey the idea clearly. 

The topological invariant we introduce in this paper can unambiguously distinguish Majorana chain from a trivial chain. Further, our argument shows that the value of the invariant must be stable. That is, for any two Hamiltonian that can be adiabatically connected to each other without closing the energy gap, the value of the invariant must be the same. Our approach is robust against both disorder and interaction, and it can be computed from a reduced density matrix of a finite-sized region.

The rest of the paper is structured as follows. In Section \ref{section:definition}, we motivate and define our topological invariant. We also compute its value for a trivial chain and Majorana chain, and compare their values. In Section \ref{section:stability}, we show that the invariant remains stable against a finite-depth quantum circuit, showing its stability against within the phase. In Section \ref{section:Renyi}, we numerically compute the invariant, corroborating the stability argument.

\section{Definition of the invariant\label{section:definition}}
 
 Recently, we showed that topological ground state degeneracy gives a rigorous lower bound to the topological entanglement entropy.\cite{Kim2013}  We show that an analogous relationship exists in one-dimensional systems as well. 

For concreteness, let us consider Kitaev's Majorana chain. Kitaev's original model can be described by the following Hamiltonian:
\begin{equation}
H=\sum_j -wa_j^{\dagger}a_{j+1} - \frac{1}{2}  \mu(a_{j}^{\dagger}a_j - \frac{1}{2}) +\Delta a_j a_{j+1}  + h.c., \label{eq:MajoranaHamiltonian}
\end{equation}
where $w$ is the hopping amplitude, $\mu$ is the chemical potential, and $\Delta$ is the superconducting gap. The operators $a_j^{\dagger}$ and $a_j$ are fermion creation and annihilation operators, and $h.c.$ is the hermitian conjugate. If $2w > |\mu|$ and $\Delta \neq 0$, a gapless boundary mode arises.\cite{Kitaev2001} Two states corresponding to each boundary modes, denoted as $\ket{0}$ and $\ket{1}$, form a set of degenerate ground states in the thermodynamic limit.

We can concoct the following global state in order to obtain an inequality between ground state entanglement entropy and ground state degeneracy:
\begin{equation}
\rho = \frac{1}{2} (\ket{0}\bra{0} + \ket{1}\bra{1}). \nonumber
\end{equation}
An important ingredient for deriving the inequality is the local indistinguishability of the two states, which asserts that they cannot be distinguished nor mapped into each other by applying any local operation.\cite{Bravyi2006}

Here, we have an interesting (weaker) variant of the local indistinguishability condition. The two states can be mapped into each other by applying a parity-violating local operation near the boundary, but the expectation values of all the local observables are identical.\footnote{If one allows only parity-conserving operation, the two states cannot be distinguished nor mapped into each other by any local operation.} Therefore, for any region that does not contain both boundaries,  $\ket{0}$ and $\ket{1}$ must be indistinguishable from each other.

Now, we partition the system into three contiguous subsystems; see FIG.\ref{FIG:InequalityChain}. Applying the strong subadditivity of entropy (SSA),\cite{Lieb1972} one can arrive at the following inequality.
\begin{equation}
S(AB) + S(BC) - S(B) \geq S(ABC) = 1. \nonumber
\end{equation}
$AB,BC,$ and $B$ are {\it local} in a sense that they have access to only half of the unpaired Majorana fermion. Therefore, all the correlation functions supported on these local subsystems must be identical for both $\ket{0}$ and $\ket{1}$. In particular, the reduced density matrices of the two states must be identical over these local subsystems. Therefore, if the system is in one of the superselection sectors, {\it i.e.}, $\ket{0}$ or $\ket{1}$ but not their superposition,  $S(AB) + S(BC) - S(B)$ must be at least larger or equal to $1$. Further, by using the purity of the states $\ket{0}$ and $\ket{1}$, we obtain the following inequality:
\begin{equation}
\gamma \geq 1, \nonumber
\end{equation}
where the {\it entropic invariant} for a one-dimensional system is defined as follows:
\begin{equation}
\gamma  := S(AB) + S(A) - S(B).
\end{equation}

There are two subtleties that are easy to gloss over. While they do not play an essential role in the main argument, an interested reader may find these facts useful. First, it is important to assume that the global state is not in a superposition of different superselection sectors. Otherwise, the entanglement entropy of some subsystem may not be equal to the entanglement entropy of its complement.\cite{Moriya2005} Second, strictly speaking, a fermionic variant of SSA must be used.\cite{Araki2003}. 

The value of $\gamma$ can be easily computed for a certain set of parameters. For example, if $w=\Delta=0$, $\gamma$ is $0$. On the other hand, if $|\Delta|=w>0$, $\mu=0$, $\gamma$ is equal to $1$.\cite{Fidkowski2010} The nontrivial lower bound on $\gamma$ suggests that these values may be stable; recall that $\gamma$ must be larger or equal to $1$ in the presence of the boundary mode. There is, of course, a remote possibility that $\gamma$ can attain a value close to $1$ for a state that can be adiabatically connected from a trivial state. We show that such a possibility is strictly forbidden. In fact, we explicitly show that $\gamma$ remains stable against any finite depth local quantum circuit, so long as $\gamma$ is {\it deformation invariant,} a concept that we shall explain in the next section.

\section{Deformation invariance and stability\label{section:stability}}

It is known that the stability of topological entanglement entropy  in two spatial dimensions can be attributed to the existence of conditionally independent subsystems.\cite{Kim2012c} Here, we use a similar idea in order to establish a stability statement for $\gamma$. 

Let us first begin by defining the key concepts. We shall colloquially refer to a linear combination of entanglement entropy to be deformation invariant if it is invariant under a small deformation of the subsystems that preserves the topology. For example, the mutual information between $A$ and $C$ is deformation invariant for a trivial state. A more interesting example is the Majorana chain. More precisely, one can set the parameters of Eq.\ref{eq:MajoranaHamiltonian} as $\Delta=w>0$ and $\mu=0$. By using the well-known formula for the entanglement spectrum, one can easily show that $I(A:C) = 1$ as long as $B$ is not an empty set.\cite{Fidkowski2010}

We shall exploit the deformation invariance by approximating the adiabatic evolution by a finite-depth local quantum circuit. Such a circuit is defined as follows:
\begin{equation}
U_{lqc} = \Pi_{i=1}^{n} U_i,\label{eq:lqc}
\end{equation}
where $n$ is a constant that is independent of the system size, and $U_i$ is a product of unitary operators which have disjoint and geometrically local supports.

\subsection{Stability argument}
We say that the quantum circuit $U_{lqc}$ has depth $n$ and width $w$ if (i) $U_{lqc}$ can be expressed as in Eq.\ref{eq:lqc} and (ii) $U_i = \otimes_j U_{i,j}$, where $U_{i,j}$ are unitary operators on disjoint supports with the support size bounded by $w$. We further assume that each of the supports are local, in a sense that they can be contained in a ball of finite radius. After a sufficient amount of coarse-graining, one can set $w$ to be $2$. Also, we define
\begin{equation}
\gamma_m^l:= I(A_l:C_l)_m, \nonumber
\end{equation}
where $A_l$ is the first $l$ sites of the lattice, $C_l$ the last $l$ sites of the lattice, and $I(A_l:C_l)_m$ is the mutual information between $A_l$ and $C_l$ for a global state $\ket{\psi_m} := \Pi_{i=1}^{m} U_m \ket{\psi_0}$.

Our argument relies on two well-known facts. First, SSA holds for general quantum states. Second, entanglement entropy of a region is invariant under a unitary operation that is supported on its region. Let us see how these two facts can be used. 

We would like to first provide an upper bound on $\gamma_m^l$. First, note the following identity:
\begin{widetext}
\begin{equation}
\gamma_m^{l+1}-\gamma_m^l = I(A_{l+1}:C_{l+1}) - I(A_{l+1}:C_{l}) + I(A_{l+1}:C_{l}) - I(A_l:C_l).
\end{equation}
\end{widetext}
Using SSA, one can see that $I(A_{l+1}:C_{l+1}) - I(A_{l+1}:C_{l})\geq 0$. A similar conclusion can be drawn for the remaining term as well. Hence, we conclude the following:
\begin{equation}
\gamma_m^l \leq \gamma_m^{l+1}.
\end{equation}
Now, we observe that the local unitary operators on the $m$th step are only supported on either $A_{l+1}$, $C_{l+1}$, or the complement of $A_{l+1}C_{l+1}$. Therefore, the entanglement entropy of these regions for $\ket{\psi_m}$ must be identical those of $\ket{\psi_{m-1}}$. The resulting upper bound is the following:
\begin{equation}
\gamma_m^{l} \leq \gamma_{m-1}^{l+1}.
\end{equation}

\begin{figure}[h]
\includegraphics[width=2.8in]{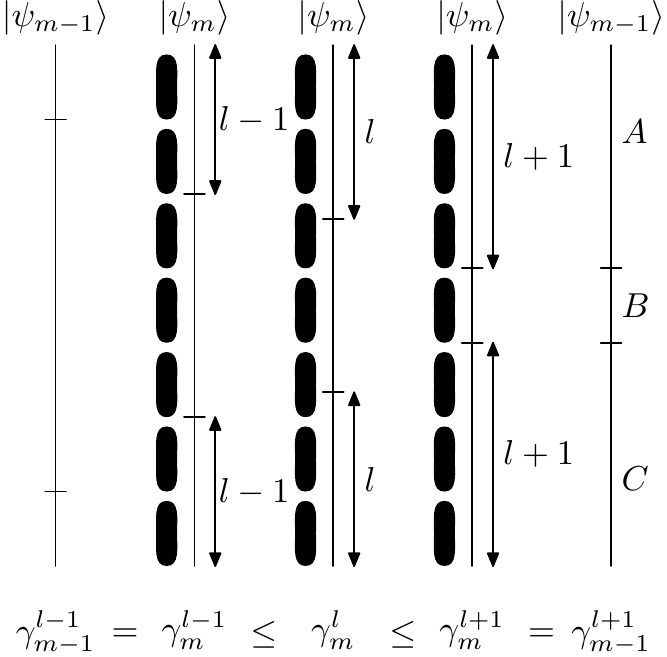}
\caption{A schematic representation of the sequence of inequalities leading to Eq.\ref{eq:recursive_inequality}. The shaded region represents a local unitary transformation $U_m$, and the vertical line represents the quantum state $\ket{\psi_{m-1}}$. Each segments represent the partition of the system into $A,B,$ and $C$. For the first and the last equality, we have used the fact that entanglement entropy is invariant under a local unitary transformation. The inequalities are simple consequences of SSA. \label{FIG:InequalityChain}}
\end{figure}

One can apply a similar argument to obtain the lower bound as well. The resulting bounds can be summarized as follows; see also FIG.\ref{FIG:InequalityChain}.
\begin{equation}
\gamma_{m-1}^{l-1} \leq \gamma_m^{l} \leq \gamma_{m-1}^{l+1}.\label{eq:recursive_inequality}
\end{equation}

Recursively applying Eq.\ref{eq:recursive_inequality}, the stability bound is obtained:
\begin{equation}
\gamma_{0}^{l-m} \leq \gamma_m^{l} \leq \gamma_{0}^{l+m}.\label{eq:sandwich_bound}
\end{equation}
Applied to a trivial state, Eq.\ref{eq:sandwich_bound} implies that $\gamma_n^{l}=0$  for $ n \leq \min(\frac{L}{2}-l, l)$.  Similarly, we conclude that $\gamma_n^l= 1$ for the Majorana chain. Therefore, if the size of the subsystems are sufficiently large, $\gamma$ should remain invariant up to a small error that decays sufficiently fast with the subsystem size. For a realistic system with a finite correlation length $\xi$, we expect the correction term to be $O(e^{-l/\xi})$, where $l$ is the size of the subsystem. 

At the transition between two phases, the entanglement should be described by the formula of Calabrese and Cardy.\cite{Calabrese2004} Due to the logarithmic dependence on the length of the subsystem, $\gamma$ will not be a universal number that is independent of the length of each subsystems.

\section{Renyi entropy analogue and its numerical benchmark\label{section:Renyi}}

Owing in part to the strong subadditivity of entropy, we have argued that $\gamma$ is an invariant that characterizes the phase. On one hand, this result is encouraging. One can compute the invariant by simply looking at the reduced density matrix of some local region that contains one of the boundaries. Therefore, if two systems have different values of $\gamma$, one can unambiguously tell that they are in a different phase. On the other hand, our result has its shortcomings. Entanglement entropy is an extremely hard quantity to measure experimentally. To the best of author's knowledge, there is no simple known way of computing the von Neumann entropy without explicitly calculating all the eigenvalues of the density matrix.

In the studies of two-dimensional systems, it has been known for a while that Renyi entanglement entropy can be used as an alternative to the von Neumann entanglement entropy.\cite{Flammia2009} Further, several proposals have been recently made to measure the Renyi-2 entanglement entropy of a generic quantum many-body system.\cite{Daley2012,Abanin2012,Pichler2013} Therefore, it is natural to ask if the Renyi entropy variant of our order parameter shows a similar behavior.

\begin{figure}[h]
\centering
\subfigure[$\eta=0$]{\includegraphics[width=0.23\textwidth]{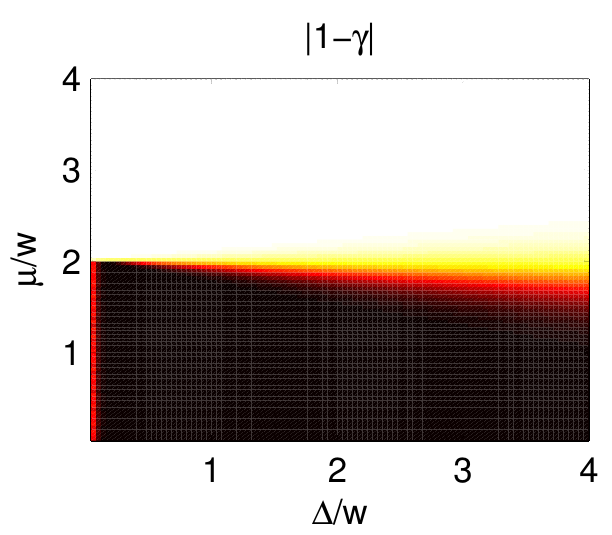}}
\subfigure[$\eta=0$]{\includegraphics[width=0.23\textwidth]{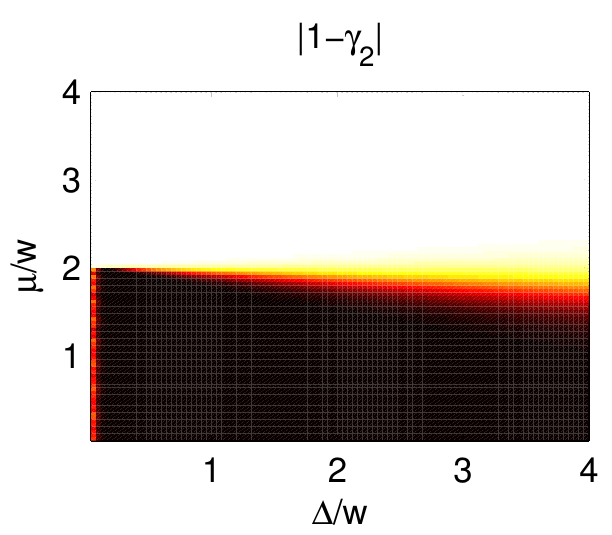}} \\
\subfigure[$\eta=0.5$]{\includegraphics[width=0.23\textwidth]{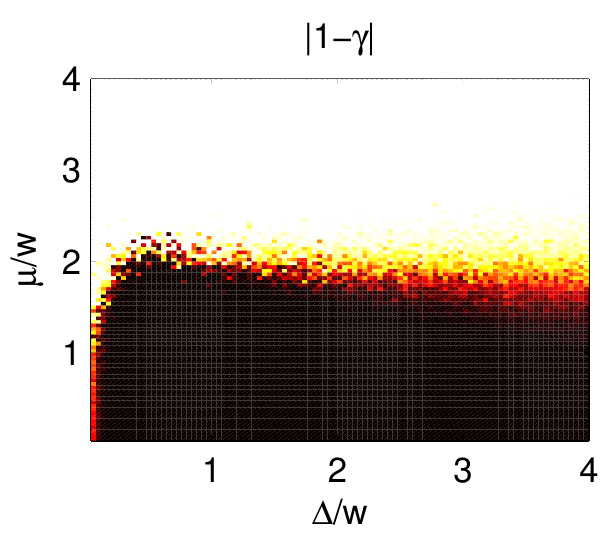}}
\subfigure[$\eta=0.5$]{\includegraphics[width=0.23\textwidth]{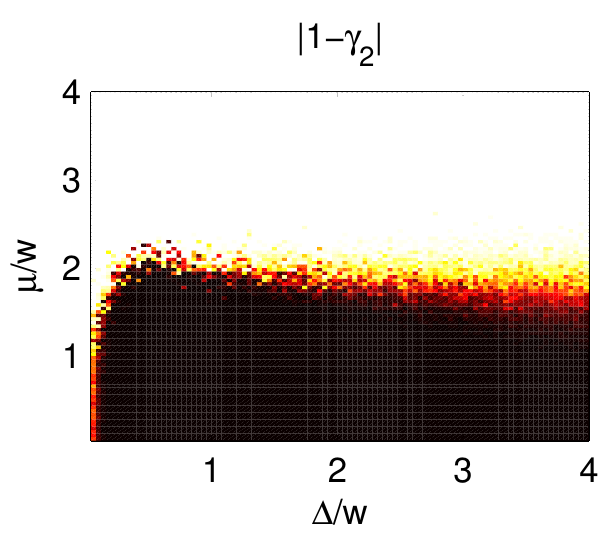}}\\
\subfigure{\includegraphics[width=0.23\textwidth]{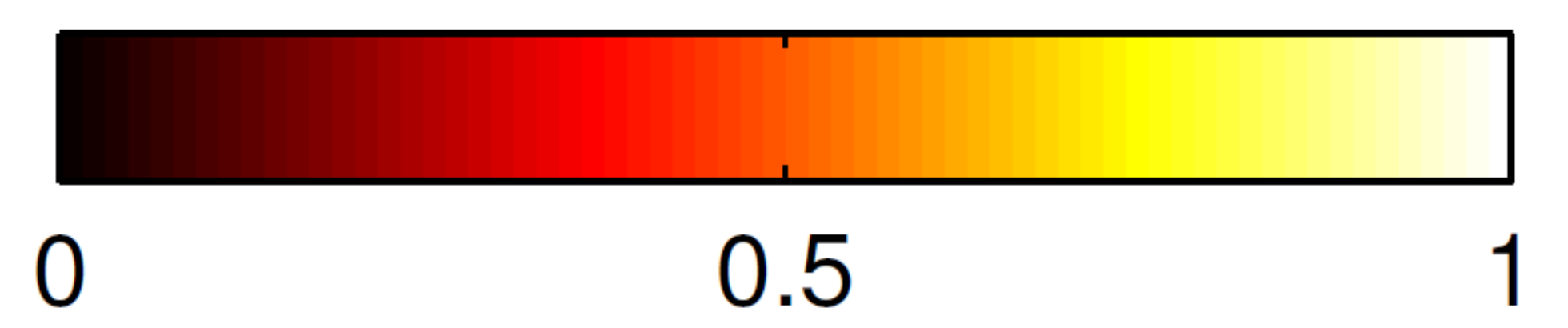}}
\caption{(color online) We have plotted the value of $|1-\gamma|$ and $|1-\gamma_2|$ over a range of parameters. The length of the chain is $60$, and the size of the subsystem $A,B,$ and $C$ were all set to $20$. We have allowed the chemical potential $\mu = \mu_0(1 + \eta x)$ to have a disorder, where $\eta$ is the disorder parameter and $x\in[-1,1]$ is a uniform probability distribution. The value of $\gamma$ and $\gamma_2$ was obtained without any averaging over the disorder probability distribution. Without disorder, {\it e.g.}, (a) and (b), the value of $\gamma$ and $\gamma_2$ both faithfully represent the phase as expected. Even in the presence of disorder, {\it e.g.}, (c) and (d), the order parameters are highly reliable except near the phase boundary. \label{FIG:benchmark}}
\end{figure}

Here, we attempt to make a similar approach by defining a Renyi-2 entropy variant of $\gamma$:
\begin{equation}
\gamma_2 := S_2(A) + S_2(AB) -S_2(B),
\end{equation}
where $S_2(\rho) = -\log \Tr(\rho^2)$. The result is plotted in FIG.\ref{FIG:benchmark} , together with the plot for $\gamma$. While the stability argument presented in this paper cannot be applied to $\gamma_2$, its behavior under the change of the parameters shows an excellent agreement with that of $\gamma$. Our numerical result indicates that $\gamma_2$ might serve as a viable candidate for unambiguously distinguishing Majorana chain from a trivial chain.

\section{Discussion}

In this paper, we have proposed a new order parameter for a generic one-dimensional system and showed that it is stable under an adiabatic evolution. While $\gamma$ is likely to be difficult to measure in a realistic experimental setting, one might be able to measure $\gamma_2$ by employing a recent proposal by Pichler {\it et al}.\cite{Pichler2013}, which is particularly well-suited for measuring the Renyi-2 entropy of optically trapped fermionic atoms. In light of their proposal, the optical realization of the Majorana chain\cite{Jiang2011} will be an interesting testbed for detecting a smoking-gun signature of the long-range entanglement. It is worth noting that the Majorana chain sidesteps the key drawback that plagues essentially all the current proposals for measuring entanglement entropy: that the accuracy of the measurement grows exponentially with the entanglement entropy. Since Majorana chain is gapped, all of its entanglement entropy must be bounded by a constant. However, it should be also noted that the stability argument given in this paper does not hold for $\gamma_2$. Therefore, more work is needed to understand its stability.

We would also like to point out that the origin of the $\Omega(1)$ deviation of the order parameter from $0$ can be traced back to (i) the existence of the states that have the same local reduced density matrices and (ii) the global superselection rule. A linear combination of the two states may lead to a different local reduced density matrix, but such a state does not follow the superselection rule. Nevertheless, for a state that satisfies the superselection rule, the invariant $\gamma$ is a well-defined quantity. In the case of the Majorana chain, the superselection rule was dictated by the global parity conservation law. It will be interesting to apply the invariant in the context of different symmetry, which is left for future work.

We would like to point out that there are other methods that achieve a similar goal. For example, Kitaev has originally proposed a Pfaffian formula which can be used for a clean, translationally invariant system.\cite{Kitaev2001} For a disordered chain, there is a prescription proposed by Akhmerov {\it et al.}\cite{Akhmerov2010} Alternatively, one may study the entanglement spectrum along a real-space cut.\cite{Fidkowski2011,Turner2010} It will be interesting to compare our approach to these previous approaches.

We close with a remark that one must be careful about the possibility that the global fermion parity is not exact. Such a scenario may occur if the environment can couple to the gapless boundary mode. However, even under such condition, the stability argument for $\gamma$ remains to be valid. Therefore, if one can observe a significant deviation of $\gamma$ from $0$ under the change of the parameters describing the chain, one may interpret it as an evidence that the chain is in a topologically nontrivial phase.

{\it -- Acknowledgement}

I would like to thank Olivier Landon-Cardinal for clearing up the initial confusion I had in the subject. I would also like to thank John Preskill, Alexei Kitaev, Sankar Das Sarma, Jay Sau, Roger Mong, and Alexey Gorshkov for the helpful discussions. This research was supported in part by NSF under Grant No. PHY-0803371, by ARO Grant No. W911NF-09-1-0442, and DOE Grant No. DE-FG03-92-ER40701.

\bibliography{bib}

\end{document}